\documentstyle[preprint,epsfig,aps]{revtex}

\draft

\newcommand{\be}{\begin{equation}}
\newcommand{\ee}{\end{equation}}

\begin{document}

\title{Comparison of the Standard Statistical Thermodynamics (SST) with 
the Generalized Statistical Thermodynamics (GST) Results for the Ising Chain}
   
\author{ U\v{g}ur TIRNAKLI\thanks{e-mail: tirnakli@fenfak.ege.edu.tr} , 
Do\v{g}an DEM\.{I}RHAN and Fevzi B\"{U}Y\"{U}KKILI\c{C} \\
Department of Physics, Faculty of Science,\\ Ege University 
35100, Bornova \.{I}zmir-TURKEY}

\date{\today}

\maketitle
   
\begin{abstract}

In this study, the internal energy and the specific heat of the
one-dimensional Ising model obtained in the frame of the generalized
statistical thermodynamics (GST) by R.F.S. Andrade (ref.[19]), have been
extended somewhat and compared with the internal energy and the specific
heat which have been calculated in the standard (conventional) statistical
thermodynamics. In the $q\rightarrow 1$ case, approaching of the generalized
internal energy and the specific heat to the standard well-known results,
have been illustrated by numerical applications. In addition to these,
$N>>1$ case has been also investigated. 

\end{abstract}

\section{Introduction}

Boltzmann-Gibbs (BG) statistics is essentially derived from the standard
Shannon entropy

\begin{eqnarray*}
S_1=-k_B\sum_{i=1}^{W} p_i \log p_i
\end{eqnarray*}

\noindent which is an extensive and concave quantity (the sub index 1 of $S$
will be defined afterwards). Although BG statistics is suitable for handling
a large number of physical systems, it fails whenever
\begin{itemize}
\item the range of the microscopic interactions is large compared to the
linear size of the macroscopic systems (long-range interactions)
\item the time range of the microscopic memory is large compared to the
observation time (non-Marcovian processes).
\end{itemize}

\noindent These kinds of violations are met for a long time in gravitation
[1], magnetic systems [2], anomalous diffusion [3] and surface tension
problems [4]. The way out from these problems seems to be Nonextensive
Statistical Thermodynamics which must be a generalization of the BG
statistics in a manner that allows a correct description of the nonextensive
physical systems as well.
The study of nonextensive formalisms is very vivid nowadays in Physics and
these formalisms keep growing along two apparently different lines:
Generalized Statistical Thermodynamics (GST) and Quantum Group-like
Approaches (QGA). Although they seem to be very different from each other,
recently it is pointed out that there exist a connection between them [5].
Here, the former one will be the subject of interest and the latter is
completely out of the scope of the present study.

GST has been proposed by C. Tsallis in 1988 [6]. The multifractal inspired
entropy which has been expressed by Tsallis in the form

\be
S_q=k \frac{1-\sum_{i=1}^{W} p_i^q}{q-1}
\ee

\noindent is a generalized entropy having the properties of positivity,
equiprobability, concavity and irreversibility. Furthermore, the standard
additivity property has also been appropriately generalized (in the $q\neq
1$ case it becomes non-extensive). In eq.(1), $k$ is a positive constant,
$q$ (Tsallis $q$-index) is a real constant associated with the number of 
measurement in the multifractal theory and $W$ is the total configuration
number. Commencing with the entropy definition of eq.(1), a canonical
distribution

\be
p_i=\frac{\left[1-\beta (q-1)\varepsilon
_i\right]^{\frac{1}{q-1}}}{Z_q}\;\;\; , \;\;\; Z_q=\sum_{i=1}^{W}
\left[1-\beta (q-1)\varepsilon _i\right]^{\frac{1}{q-1}}
\ee

\noindent  could be obtained for a system in a heat bath, by maximizing the
entropy under suitable constraints [6].

Soon after Tsallis' work, the generalization of the conventional concepts in
various fields of Physics have rapidly proceeded. Among the successfully
tackled concepts in the frame of this generalization, generalized
thermostatistics [7], Boltzmann $H$-theorem [8,9], Ehrenfest theorem [10],
the distribution function of the classical and quantum gases [11,12],
Langevin and Focker-Planck equations [13], Bogolyubov inequality [14],
fluctuation-dissipation theorem [15], von Neumann equation [16], classical
equipartition principle [17], Callen identity [18], Ising chain [19,20], 
paramagnetic systems [21], infinite-range spin-$1/2$ Ising ferromagnet [22], 
mean-field Ising model [23], the generalized transmissivity for spin-$1/2$ 
Ising ferromagnet [24], self-dual planar Ising ferromagnet [25] and Planck
radiation law [26,27] could be enumerated.

On the other hand, this generalization has also been succsessfully used to
overcome the failure of BG statistics in some physical applications such as
stellar polytropes [28], the specific heat of the unionized hydrogen atom
[29], Levy-like anomalous diffusion [30], $d=2$ Euler turbulence [31], solar
neutrino problem [32] and velocity distribution of galaxy clusters [33].

Up to now, a very large number of works, which have been performed in the
frame of GST, are devoted to the investigation of the magnetic systems.

As an example of these, the distribution function of the paramagnetic
systems might be noted. In this study, for the paramagnetic systems
exhibiting multifractal structures, the connection between the Tsallis
$q$-index and the fractal dimension and the range of the values of the
$q$-index have been established [21].

In another work [23], using the generalized Bogolyubov inequality, the free
energy and the magnetization of mean-field Ising model have been obtained
within GST and it is noticed that the range of the $q$ values are consistent
with the one obtained in ref.[21]. Moreover, the expression obtained for the
critical temperature is exactly the same as the one found in ref.[18].

Another example for the investigation of the magnetic systems in the frame
of GST is the Ising chain which is the subject of the present work. In his
first study relevant to this subject, Andrade calculated [19] the internal
energy and the specific heat of the Ising chain by making use of the
distribution given in eq.(2) which had been given before by Tsallis. Later
on, in his second investigation [20] he reiterated his calculations using
the distribution

\be
p_i=\frac{\left[1-\beta (1-q)\varepsilon
_i\right]^{\frac{1}{1-q}}}{Z_q}\;\;\; , 
\ee 
 
\noindent where the partition function, $Z_q$ is

\be
 Z_q=\sum_{i=1}^{W}
\left[1-\beta (1-q)\varepsilon _i\right]^{\frac{1}{1-q}}\;\; .
\ee

\noindent At first sight, this distribution seems to be very similar to
eq.(2), only with a small redefinition of $q$, however, this is not the case
since this change originates from the fact that in the definition of the
second constraint used for maximizing the Tsallis entropy, $q$-expectation
values have been used.

Here, our motivation is to improve Andrade's first work [19] somewhat by 
considering $r=3$ case and to strengten Andrade's paper by comparing the 
obtained results with the standard ($q=1$) ones. In the present study, the
generalized internal energy and specific heat ($q\neq 1$) have been
compared with the standard results. In the $q\rightarrow 1$ case, the
transformation of the results obtained by GST to the standard results, have
been observed. Moreover, by taking into account the $N>>1$ case, the
approaching of the internal energy expressions to a constant value and thus
vanishing of the specific heat has also been illustrated. In this manner,
the present study supports the investigations of Andrade as well as
emphasizes the fact that GST is one of the most favourable tools for the
examination of the multifractal structures. (It is obvious that, the other
studt of Andrade [20] could also be tackled in the same manner.)

\section{One-dimensional Ising Model}

In the model established by E. Ising for the first time with the objective
of explaining ferromagnetism [34], a two-valued spin variable $s_i$ taking
the values of $+1$ or $-1$ has been allocated for each lattice site. The
model Hamiltonian in one-dimension, in the absence of the magnetic field,
could be written in the following form

\be
{\cal H}=-J \sum_{i=1}^{N}s_i s_{i+1}
\ee

\noindent where $J$ is the exchange parameter and $N$ is the number of total
lattice sites.

The analytical solution of the model with the transfer matrix method of
standard statistical thermodynamics (SST) using the Hamiltonian of eq.(5)
leads to the following expressions for the internal energy and the specific
heat respectively [35] :

\be
\frac{E}{J}=-N\tanh(\beta J)
\ee

\noindent and

\be
\frac{C}{k_B}=N\beta ^2J^2\sec h^2(\beta J) \;\;\; .
\ee

\section{Obtaining Some of the Thermodynamic Quantities of the Ising Chain
Using GST}

The most prominent difficulty met in the solution of systems investigated
with GST is the determination of the partition function. This situation
originates from the non-extensivity property which arises for the case
$q\neq 1$. Andrade wrote the Hamiltonian of eq.(5) with a parameter $\bar{E}$
which calibrates the ground state energy of the system for the Ising chain
in the following form

\be
{\cal H}=\bar{E}-J\sum_{i=1}^{N}s_i s_{i+1}
\ee

\noindent and by taking $\bar{E}=-NJ$ such that the values of the energy
levels would become nonpositive from $-2NJ$ to $0$, he expressed the
partition function of the system in a solvable form within the frame of GST
by combinatorial method. As a consequence, he established the following
expression [19] :

\be
E_r=-NJ-J \frac{\sum_{m=0}^{r}\sum_{s=0}^{r-m}\left(\frac{\beta
J}{r}\right)^{r-s}\frac{r!}{s!m!\left(r-m-s\right)!}\sum_{u=0}^{m+1}
\zeta_{m+1,u}N^{r+u-m-s}}{\sum_{m=0}^{r}\sum_{s=0}^{r-m}\left(\frac{\beta
J}{r}\right)^{r-s}\frac{r!}{s!m!\left(r-m-s\right)!}\sum_{u=0}^{m}
\zeta_{m,u}N^{r+u-m-s}}
\ee

\noindent where $r=1/(q-1)$.

By writing down the definitions 

\be
\zeta_{mu}=\sum_{n=0}^{u}C_m(n)\gamma_{m-n,u-n}(-1)^n
\ee

\be
C_m(n)=\frac{m!}{n!(m-n)!}
\ee

\be
\gamma_{mn}=\sum_{s=0}^{m}\eta_{ns}^{m}\;\;\;\;\;(m,n,s\geq 0)
\ee

\be
\eta_{ns}^{m+1}=\eta_{n-1,s-1}^{m}+(s-m-1)\eta_{n,s-1}^{m}+s\eta_{ns}^{m}
\ee

\noindent the internal energy values for $r=1 (q=2), r=2 (q=1.5), r=3
(q=1.33)$ from eq.(9) could be written down as :

\be
E_1=-NJ-\frac{\beta J^2N}{\beta JN+1}
\ee

\be
E_2=-NJ-\frac{\beta J^2N\left(\frac{1}{2}\beta JN+1\right)}{\frac{1}{4}\beta
JN\left[\beta J(N+1)\right]+1}
\ee

\be
E_3=-NJ-\frac{\beta
J^2N\left[\frac{\beta^2J^2}{9}\left(N^2+N-\frac{2}{3}\right)+\frac{2}{3}\beta
JN+1\right]}{\frac{1}{3}\beta
JN\left[\frac{\beta^2J^2}{3}\left(N+\frac{N^2}{3}\right)+\beta
J(N+1)\right]+1}\;\; .
\ee

\noindent The variations of the internal energy with respect to the reduced
temperature $(1/\beta J)$ for eqs.(14)-(16) of GST and eq.(6) of SST have
been illustrated in Fig.1. In accordance with the expectations, for
increasing $r$ values ($q$ values decreasing to 1) the behaviour of the
internal energy curve approaches to that of the analytical solution of SST.
Furthermore, in $N>>1$ case, as it is illustrated in Fig.2, $E_r$ approaches
to a constant value which is an unexpected result.

From the internal energy expressions given by eqs.(14), (15) and (16),
taking $r=1 (q=2), r=2 (q=1.5), r=3 (q=1.33)$, the following conclusions
could respectively be obtained for the specific heat :

\be
\frac{C_1}{k}=\frac{\beta J^2}{\left(1+\beta JN\right)^2}
\ee

\be
\frac{C_2}{k}=\frac{4\beta^2J^2\left[4+4\beta JN+\beta^2J^2N(N-1)\right]}
{\left[4+4\beta JN+\beta^2J^2N(N+1)\right]^2}
\ee

\be
\frac{C_3}{k}=\frac{9\beta^2J^2\left[\beta^4J^4\left(N^4+N^2-2N\right)-6
\beta^3J^3\left(N^3+3N^2\right)-18\beta^2J^2+108\beta JN+81\right]}
{\left[\beta^3J^3\left(N^3+3N^2\right)+9\beta^2J^2\left(N^2+N\right)+27
\right]}\;\; .
\ee

The variation of the specific heat with respect to the reduced temperature
for $N=1$ has been illustrated in Fig.3. It has been noticed that, the
curves plotted using eqs.(17),(18) and (19) of GST for $r=1,2$ and $3$
approaches to the curve given by eq.(7) for increasing $r$ values (while $q$
values decreasing to 1), which has been established by the transfer matrix
method of SST. Furthermore, in $N>>1$ case, approaching of $E_r$ to a
constant value naturally causes $C_r$ to vanish.

\section{Conclusions}

In this study, the expressions that have been obtained in GST for the
internal energy and the specific heat of the Ising chain have been compared
with the corresponding relations in SST. $q$ could be regarded as a
parameter belonging to the internal energy and the specific heat. It has
been clearly shown by numerical applications that; the internal energy and
the specific heat established for the Ising chain in the frame of GST
approach to the conventional results obtained by the Boltzmann-Gibbs
distribution in the $q\rightarrow 1$ (increasing $r$) case. This conclusion
conforms our expectations such as the transformation of the Tsallis entropy
(eq.(1)) to the Shannon entropy. The parametric effect of $q$ has been
brought to a concrete basis with the aid of the plots. The verification of
the internal energy and the specific heat expressions brought in by Andrade
have been supported by the standard results obtained by the transfer matrix
method. In addition to these, the $N>>1$ case has also been considered; the
explicit approaching of the internal energy to a constant value which inturn
leads to the vanishing of the specific heat has been graphically
illustrated.

\section*{Acknowlegment}

The authors would like to thank Prof. C. Tsallis for his helpful comments.

\newpage

\section*{Figure Captions}

\vspace{2cm}

\noindent
{\bf Figure 1.} $\;$ The variation of the internal energy with the reduced
temperature for $N=1$.

\noindent
{\bf Figure 2.} $\;$ The variation of the internal energy with the reduced
temperature for $N>>1$, namely $N=30$.

\noindent
{\bf Figure 3.} $\;$ The variation of the specific heat with the reduced 
temperature for $N=1$.

\newpage

\begin{figure}[htbp]
\begin{center}
\addvspace{1cm}
\leavevmode
\epsffile{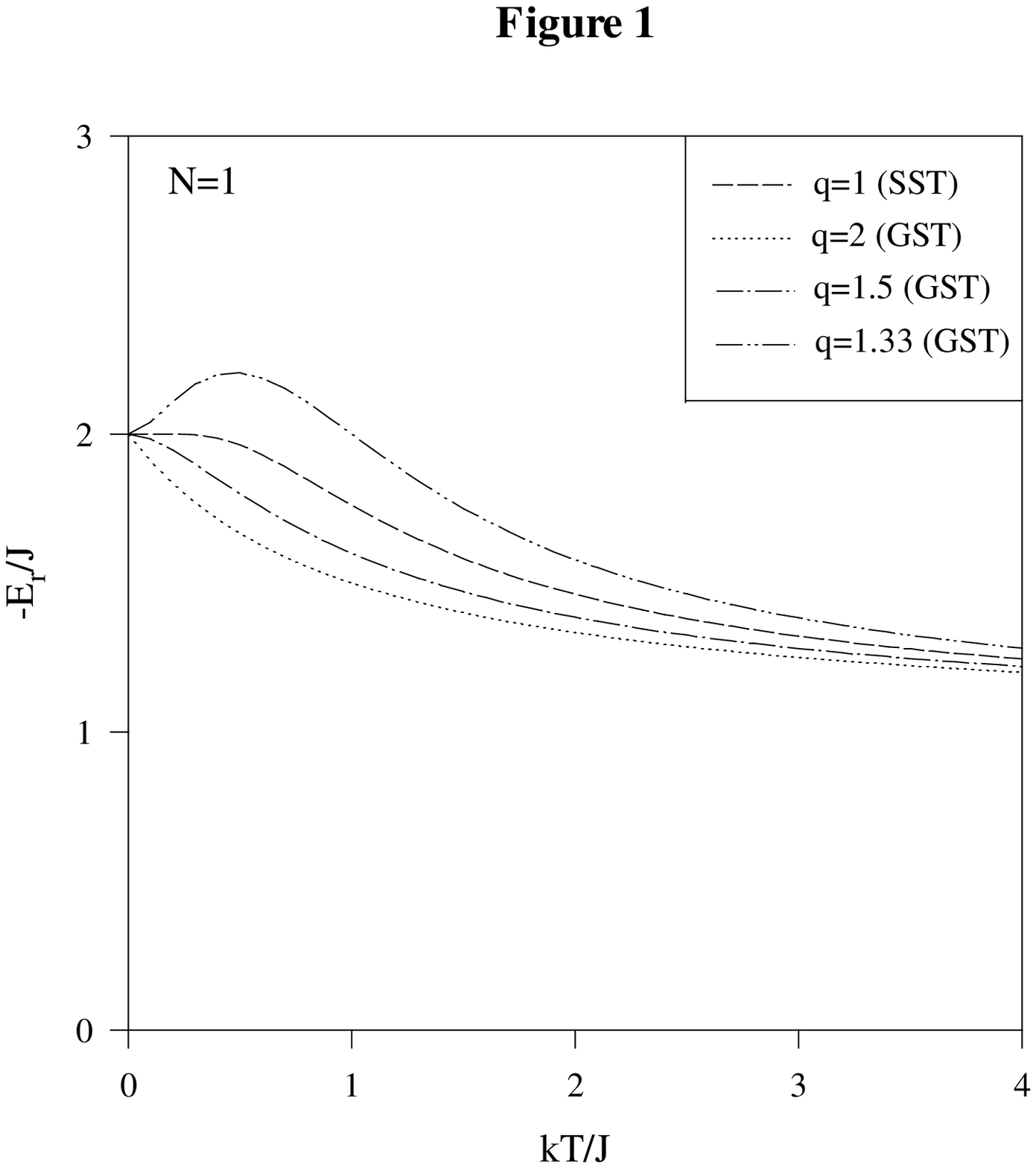}
\end{center}
\end{figure}

\newpage

\begin{figure}[htbp]
\begin{center}
\addvspace{1cm}
\leavevmode
\epsffile{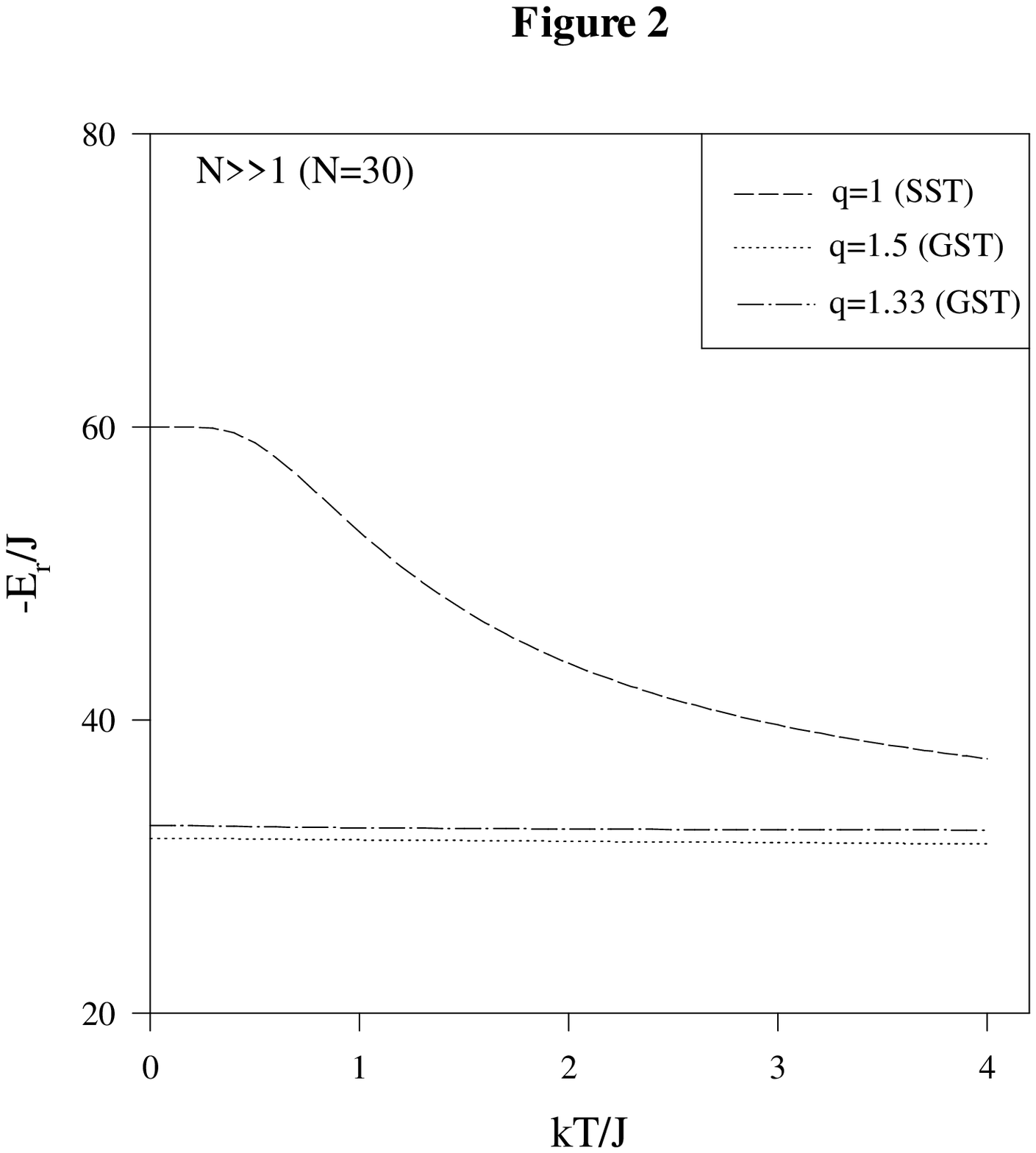}
\end{center}
\end{figure}

\newpage

\begin{figure}[htbp]
\begin{center}
\addvspace{1cm}
\leavevmode
\epsffile{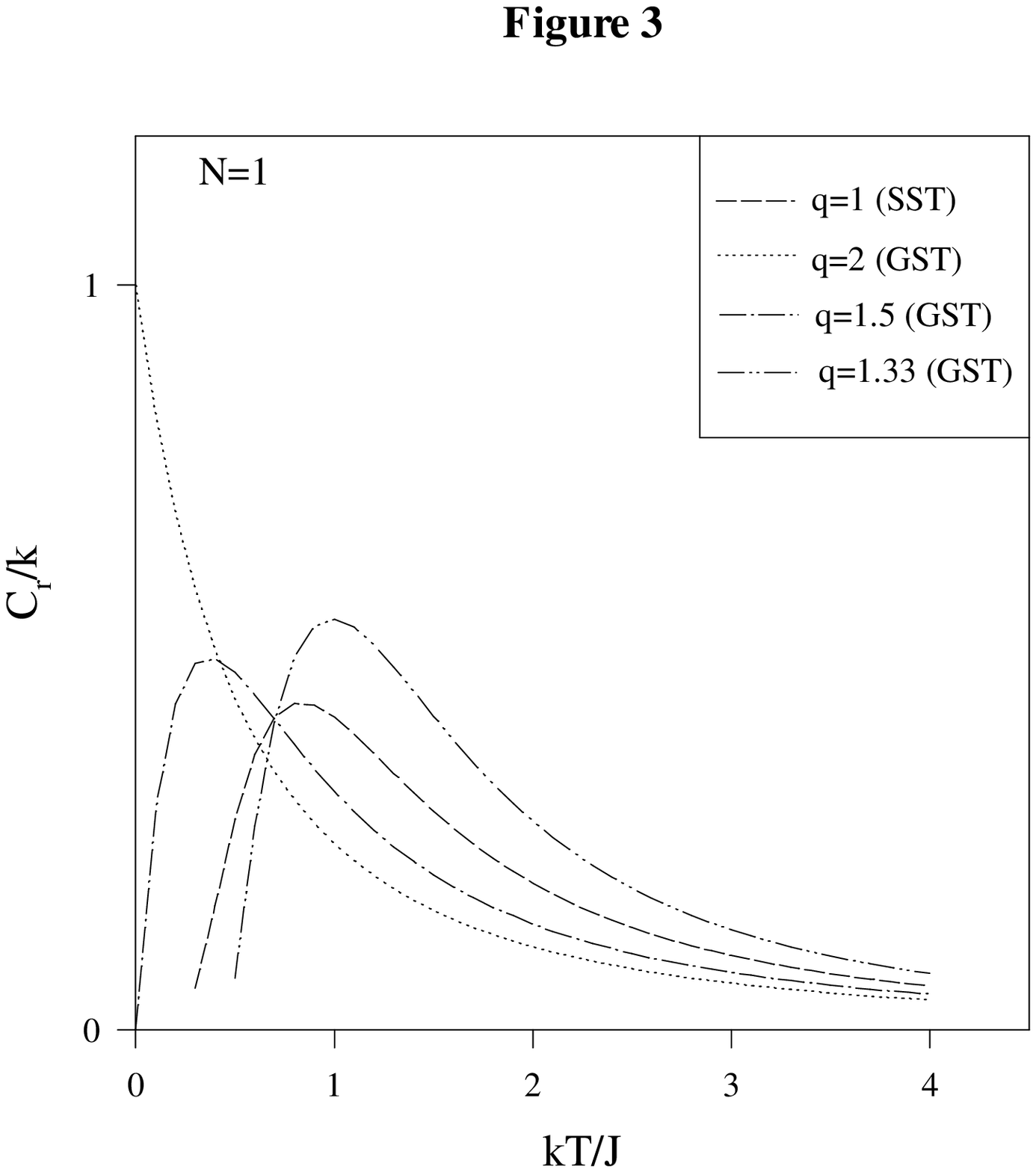}
\end{center}
\end{figure}

\end{document}